\documentclass[journal]{IEEEtran}

\usepackage{cite}
\usepackage{amsmath,amssymb,amsfonts}
\usepackage{algorithmic}
\usepackage{graphicx}
\usepackage{textcomp}
\usepackage{xcolor}
\usepackage{cleveref}
\usepackage{xcolor,colortbl}
\usepackage{lscape}
\usepackage{array}
\usepackage{makecell}
\usepackage{cellspace}
\usepackage{longtable}
\begin{document}

\title{Data-driven Model Predictive and Reinforcement Learning Based Control for Building Energy Management: a Survey}

\author{
Huiliang Zhang$^{*}$\footnote{Equal Contribution}, 
Sayani Seal$^{*}$, 
Di Wu, Benoit Boulet, Fran\c{c}ois Bouffard, \and Geza Joos\\
McGill University\\
\{huiliang.zhang2, sayani.seal, di.wu5\}@mail.mcgill.ca,
\{benoit.boulet, francois.bouffard, geza.joos\}@mcgill.ca
\vspace{-2em}}

\maketitle


\begin{abstract}
Building energy management is one of the core problems in modern power grids to reduce energy consumption while ensuring occupants' comfort. However, the building energy management system (BEMS) is now facing more challenges and uncertainties with the increasing penetration of renewables and complicated interactions between humans and buildings. Classical model predictive control (MPC) has shown its capacity to reduce building energy consumption, but it suffers from labor-intensive modelling and complex on-line control optimization. Recently, with the growing accessibility to the building control and automation data, data-driven solutions have attracted more research interest. This paper presents a compact review of the recent advances in data-driven MPC and reinforcement learning based control methods for BEMS. The main challenges in these approaches and insights on the selection of a control method are discussed.
\end{abstract}


\section{Introduction}
\label{sec_intro}

The building sector accounts for about 76\% of electricity use,
40\% of primary energy use and associated greenhouse gas emissions in the U.S. \cite{ass2015}, and the similar situation exists in other countries. Therefore, it is essential to reduce energy consumption in buildings to meet national energy and environmental challenges, and to reduce operation costs for buildings. %Hence, improving the efficiency of building energy management systems (BEMS) are significant. 
Recently, the area of building energy management system (BEMS) has gained a significant amount of interest, and the advanced control strategies for BEMS are believed to provide great potential to reduce building energy costs and improve grid energy efficiency and stability \cite{missaoui2014managing}.
%A persistent motivation for continued development in building energy management is the fact that 40\% of the total global energy consumption \cite{UN_Env} is attributed to the building sector. Over the years the scenario has not changed significantly. 
There are two main objectives for BEMS: minimizing energy consumption and ensuring the comfort of the occupants. Comfort management includes the control of the multiple components of heating, ventilation and air conditioning (HVAC) units with optimized energy consumption. Energy management deals with energy cost optimization by curtailing redundant energy usage and load shifting. However, the design of such a system faces increasing challenges and uncertainties with the penetration of renewables in the grid, which
% for sustainable and less carbon-intensive energy system \cite{IEA_renewables}, 
entails the efficient management of stationary and mobile (e.g., electric vehicle (EV)) battery storage units.
% with the increasing penetration of heating ventilation and air conditioning (HVAC), batteries, renewable energy generation and electric vehicles (EVs), the system uncertainties increase. 
These boil down to a large-scale complex optimization problem affected by multiple external disturbances and system uncertainties. Plus, these challenges complicate even more with variable occupancy conditions and human interaction with the building \cite{mariano2020review}.  

Traditional rule-based control (RBC) is easy to implement while its control performance is not optimal.
Classical model predictive control (MPC) uses forecast information to optimize the control inputs and thereby utilizing disturbance predictions for the modelling process, resulting in a reliable control performance. %MPC has thus been the state of the art for building comfort and energy optimizations over the past several years \cite{oldewurtel2012use,oldewurtel2010energy}.
% and has consistently shown to deliver improved performance \cite{COTRUFO2020109563} as compared to conventional rule based control (RBC) or proportional-integral-derivative (PID) control. 
However, several studies have pointed out its limited commercial implementation \cite{DRGONA2018199,chen2019gnu}. 
% \textit{Yang et al.} have listed recent works on real-time implementation of MPC driven HVAC control in \cite{YANG2020115147}. 
This limited adoption of MPC is attributed to predictive model design complexities, as well as increased time constraint and memory footprint required for on-line optimization \cite{cigler2013beyond}. 
The dependency of the MPC performance on the accuracy of the derived system modeling exponentially increases the computational complexity with the size of the building and the structure of its energy network.

With the growing number of smart meters and sensors installed, building operation data have recently been more accessible through the building automation systems (BAS) \cite{MADDALENA2020104211}. Empowered by big data and powerful computing resources,
%Although conventional control setups are commonly installed, this data can be efficiently used to learn the system dynamics which are essential for model-based advanced control approaches. 
a recent approach is to combine various machine learning (ML) tools with classical MPC to design data-driven MPC strategies that preserve the reliability of classical MPC while reducing the time and computational complexities during on-line implementation. 
Another trend, reinforcement learning (RL), with an excellent decision-making capability in the family of ML, is also being explored in BEMS to enhance control performance. The aim is to address consistent implementation issues, namely, time expensive on-line optimization for advanced control strategies. % \cite{kathirgamanathan2019feature,FERREIRA2012238}. 

Both of data-driven MPC and RL in BEMS are effective solutions in BEMS, and may even complement each other with their advantages. Several research efforts have been made to further understand and summarize the use of them. The authors in \cite{MADDALENA2020104211} explore the challenges of data-driven algorithms in building HVAC and energy management applications. An extensive review presented in \cite{KATHIRGAMANATHAN2021110120} highlights system model development focusing on building passive thermal mass and data-driven predictive control integration in similar building applications. A recent review \cite{yu2020deep} focuses on deep RL-based smart building energy management.  To the best of the authors’ knowledge, no detailed study has been conducted to give insights into the commonalities and interconnections between these two different control methods.  Therefore, with the aim of filling this vital research gap in this area, the main contributions of this survey are:

%Although a few related review works have been presented recently, none of them give insights into the commonalities between different control methods such as MPC and RL in BEMS.
% none of them give a whole view of research insight between MPC and RL in BEMS.  
%The authors in \cite{MADDALENA2020104211} explore the challenges of data-driven algorithms in building HVAC and energy management applications, and then present a survey of RBC, MPC, learning MPC and RL techniques. An extensive review presented in \cite{KATHIRGAMANATHAN2021110120} highlights system model development focusing on building passive thermal mass and data-driven predictive control integration in similar building applications. A recent review \cite{yu2020deep} focuses on deep RL-based smart building energy management. %However, none of them give insights into the commonalities between MPC and RL in BEMS. In this work, the interconnections between MPC, model-based data-driven MPC and RL-based algorithms in BEMS are discussed, from an application perspective.

\begin{itemize}
    \item To present a summarized review of recent advances in data-driven MPC and RL-based control methods in BEMS.
    \item To identify the main challenges and compare the difference of these two methods for the BEMS application.
    \item To provide an insight on pros and cons of each methods and suggest potential directions for engineers and researchers contemplating the use of these methods based on specific control objectives.
    
    %\item Review recent advances in data-driven MPC and RL-based control methods in BEMS,
    %\item Identify the main challenges of these two methods for the BEMS application,
    %\item Suggest potential directions for engineers and researchers contemplating the use of these methods based on specific control objectives.
    % \item Present a rough comparison, based on crucial implementation parameters, that may serve as a quick reference for engineers and researchers to select one of these methods based on specific BEMS control objectives.
\end{itemize}

 The rest of the article is organized as described in the following. A brief discussion on the technical background of MPC and RL-based control mechanism along with the building energy management problem in \Cref{sec_background}. \Cref{sec_MPC_survey} and \Cref{sec_RL_survey} include reviews of articles focusing on data-driven MPC and RL-based building energy management approaches. In \cref{sec_challenges} challenges of these control approaches are discussed followed by a discussion on future research directions in \Cref{sec_conclusion}.
\section{Technical Background} 
\label{sec_background}

In this section, a quick review of the theoretical backgrounds of MPC and RL-based building control strategies is presented.

\subsection{Model Predictive Control}
\label{subsec_MPC}

\begin{figure}[!ht]
  \centering
%  \hspace*{-0.3in}
% \captionsetup{justification=centering}
  \includegraphics[width=0.85\columnwidth]{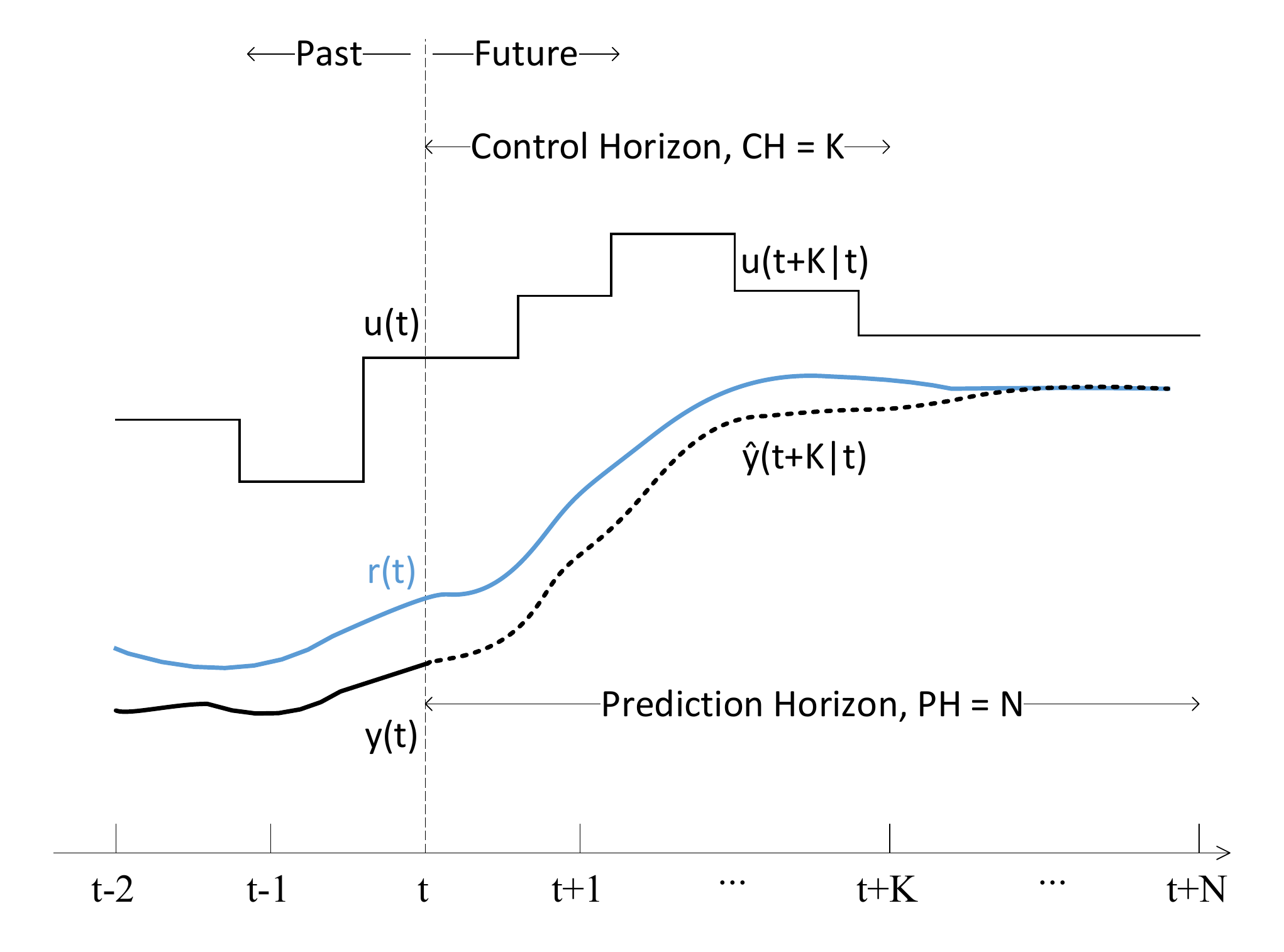}
  \caption{Model Predictive Control Strategy \cite{camacho2013model}.}
  \label{fig_C2_MPC}
\end{figure}

A schematic of a classical MPC approach is shown in \Cref{fig_C2_MPC} \cite{camacho2013model} for a tracking control problem. The MPC uses a control-oriented system model to generate output predictions, $\hat{y}(\cdot | t)$ for a predefined $N$ time steps in the future, known as the \textit{prediction horizon ($PH$)}, using the available input-output information at the current time step $t$. The controller then optimizes the manipulated input variables for a shorter interval $K$, known as the \textit{control horizon ($CH$)}, keeping the control inputs $u(\cdot|t)$ unchanged during the time interval between $t+K+1$ and $t+N$, i.e., $\Delta u(\tau | t) = u(\tau|t) - u(\tau-1|t) = 0, \ \forall \tau = t+K+1, \hdots, t+N$. The objective is to minimize a problem specific objective function based on the predicted controlled outputs. An example of a quadratic MPC objective function tracking a reference trajectory $r(t)$ is given by \Cref{equ_MPCopt}:

\vspace{-0.2in}
\begin{equation}
\min_{u(t+1| t), \hdots, u(t+K | t)} \sum_{i=t+1}^{N} w_{y} [r(i) - \hat{y}(i)]^2 + w_{u} [u(i) - u(i-1)]^2
\label{equ_MPCopt}
\end{equation}
where $w_{y}$ and $w_{u}$ are weighting coefficients. Conventionally, only the first value of the optimized control signal, $u(t+1|t)$ is implemented and the whole optimization is repeated for the next time step by shifting the prediction horizon forward, based on a new set of available information. For this reason, MPC is also known as the \textit{receding horizon control} \cite{camacho2013model}.

The design of the control-oriented system model is not trivial for building applications and the computational complexity grows with the size of the building. Also, the predictive control optimization at each time step, in a receding horizon setup, is computationally heavy and the order increases exponentially with the length of the prediction horizon as well as the number of control inputs in the optimization problem. The data-driven MPC incorporates ML-based algorithms to navigate through these computational complexities focusing on reducing time and memory footprints acquired by a classical MPC. As discussed later in \Cref{sec_MPC_survey}, generally in data-driven MPC, the ML algorithms are used either to imitate the control action of a classical MPC during run-time to reduce the online optimization time or to train an ML network to represent the control oriented building model using historical building operation data. The latter may also help in improving the adaptability of a data-driven MPC approach as compared to a classical MPC.

%\vspace*{-0.3in}

% MPC fits for the building comfort and energy management applications as it gives a suitable setup to predict the thermal comfort and energy expenses based on the available weather forecast, predictions of occupancy conditions, time-of-use electricity rates (TOU), forecast for energy market pricing etc. and thereby generate optimal control strategy by solving a single constrained multi-objective cost function \cite{privara2011role,perera2017comparison,paris2010heating,hazyuk2012optimal_part1,afram2014theory}.

%%%%%%%%%%%%%%%%%%%%%%%%%%%%%%%%%%%%%%%%%%%%%%%%%%%%%%%%%%%%%%%%%%%%%%%%%%%%%%%%%%%%%%%%%%%%%%%%%%%%%%%%%%%%%%%%%%%%%%%%%%%

\subsection{Reinforcement Learning}

\begin{figure}[h]
\centering
\graphicspath{{figures/}}
\includegraphics[width=7.1cm,height=2.7cm]{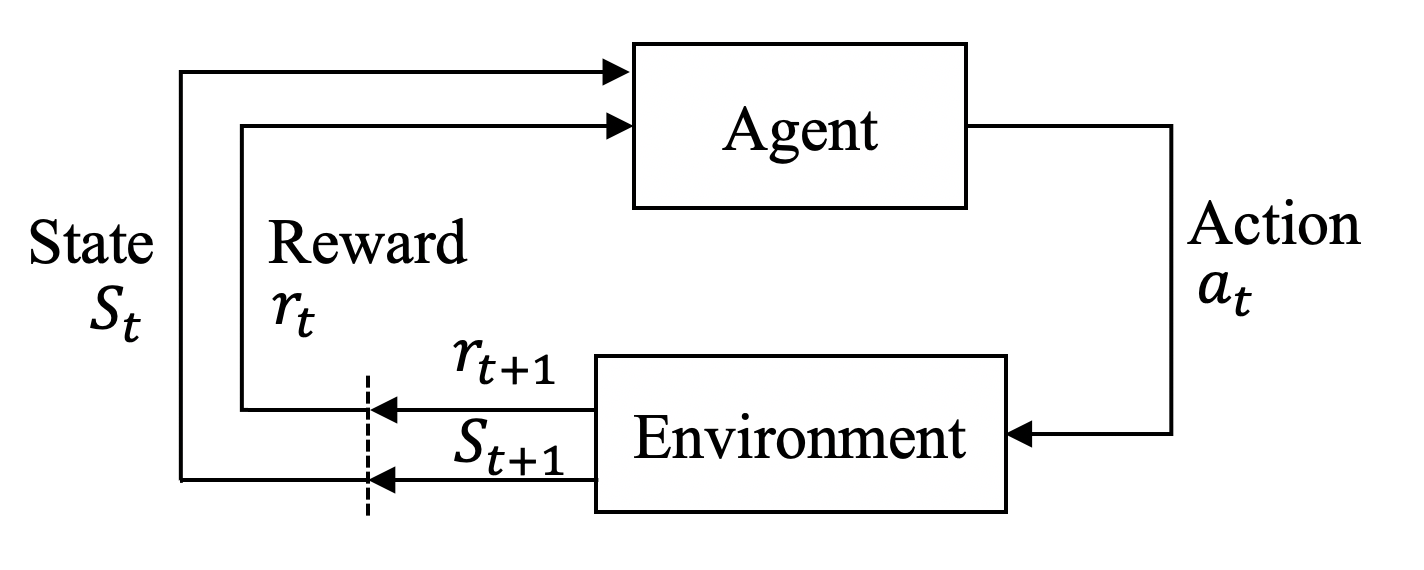}
\caption{RL Diagram.}
\label{fig_RL}
\end{figure}

RL refers to a computational approach to learning whereby an agent tries to maximize the total amount of reward it receives while interacting with a complex and uncertain environment,
%The agent acts as a learner and a decision-maker. The environment, in return, provides rewards and a new state based on the actions of the agent, 
as shown in \Cref{fig_RL}. The most common way to model an RL problem is as a Markov Decision Process (MDP). MDP is a discrete-time framework for modelling multi-stage decision making. It can be expressed as a tuple $\left<S,A,P,R,\gamma\right>$, where $S$, $A$, $P$ and $R$ are the sets of states $s_t$, actions $a_t$, state transition probabilities $p$ and rewards $r$; $\gamma\in[0,1]$ is a discount factor accounting for future rewards.
\iffalse
The mathematical model of RL can be formulated as follows:
\begin{eqnarray}\label{e1}
{p_\theta(s_1,a_1,...,s_T,a_T)=p(s_1)\prod_{t=1}^{T}\pi_\theta(a_t|s_t)p(s_{t+1}|s_t,a_t),}
\end{eqnarray}
\begin{eqnarray}\label{e2}
{\max \limits_{\theta}{\left\{E_{\Gamma\sim{p_\theta(\Gamma})}\left[\sum_t{\gamma*r(s_t,a_t)}\right]\right\}}},
\end{eqnarray}
where the sequence of states and actions $s_1,a_1,...,s_T,a_T$ is called trajectory $\Gamma$, which is determined by the transition probability $p(s_{t+1}|s_t,a_t)$ and the optimal policy $\pi_\theta(a_t|s_t)$. The expectation operator implies both the environment and the policy could be stochastic. 
\fi
RL algorithms can be sorted into model-based and model-free methods based on whether $P$ and $R$ are learned first (model-based) and then used to find the optimal control policy. 
%The model-free RL explores the optimal policy via trials and error from the interactions with the environment directly, which skips the process of having to learn the model. 
There are also three types of methods based on what the agent learns in model-free RL: value-based, policy-based and Actor-Critic methods.  The State-Action-Reward-State-Action (SARSA) and Q-learning are the most famous classical RL algorithms, which are value-based. Policy-based methods learn the policy directly with a parameterized
function respect to $\theta$. To overcome the high variance in gradient estimates, the Actor-Critic methods parameterize both policy and value functions and simultaneously update them in training. To scale up the RL algorithms with large state and action spaces, deep RL leverages deep neural networks as function approximators to represent policies, reward and value functions. The use of replay buffer and fixed target network also fix the problems of correlations between samples and non-stationary targets in classical RL algorithms \cite{mnih2015human,lillicrap2015continuous}. 

\subsection{Architecture for BEMS}
\begin{figure}[h]
\centering
\graphicspath{{figures/}}
\includegraphics[width=1.02\columnwidth]{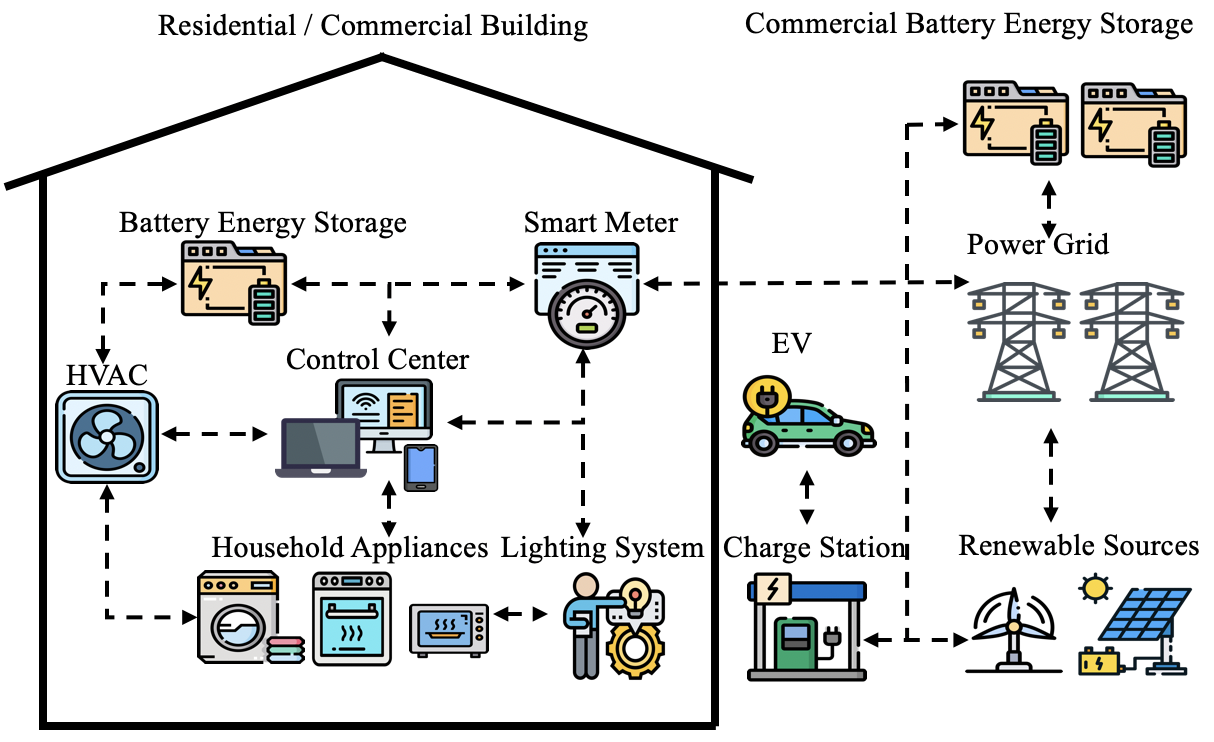}
\caption{Architecture for BEMS.}
\label{fig_ArchTec}
\end{figure}

A typical BEMS architecture has several important components: residential or commercial buildings, EVs, battery energy storage, renewable sources and the power grid, as illustrated in \Cref{fig_ArchTec}. The building itself contains household or commercial appliances and equipment, with a smart meter and the control center used to record data and control the equipment in the building. The information flow between the assets and control center is two-way to keep the building working and share information with the external power grid. 
%The HEMS control center often consists of five major functions like monitoring, logging, control, management and alarm. 

Building operation goals commonly deal with the trade-off of ensuring occupants' comfort while minimizing the energy consumption or the associated cost.  The metrics for energy optimization mainly focus on net energy consumption such as cost minimization, peak reduction, load shifting, and appliance scheduling. The comfort optimization metrics focus on keeping control variables like indoor temperature and humidity within a reasonable range.
%The metric of the controller performance varies with the design of desired control mechanism. For thermal comfort optimization primarily follows directives from ASHRAE standards \cite{Ashrae}. Two typical types of comfort optimization metrics are: (1) maintaining air temperature of a thermal space within the user-defined high and low temperature set points \cite{}; a variation of this is optimizing operative temperature \cite{} (a weighted sum of the air temperature and mean radiant temperature) instead of the air temperature; (2) maintaining predicted mean vote (PMV) index to calculate thermal sensation based on occupants' votes on a predefined point scale, and predicted percentage dissatisfied (PPD) index a function of PMV \cite{Eini_2020}. PMV is a nonlinear function of human metabolism, human external activity rate, human clothing factor, air velocity, air humidity, air temperature and radiant temperature \cite{Eini_2020,fanger1970thermal}. The metrics for energy optimization, on the other hand, may focus on net energy consumption and cost minimization, peak reduction, load shifting, appliance scheduling etc.

%!TEX root = AI_MPC_buildingCOntrol_survey.tex

\section{Review of Data-driven MPC in BEMS}
\label{sec_MPC_survey}

% One of the earlier implementations of neural network (NN) based MPC is presented in \cite{NIPS1999_db957c62} in the process industry. Here, the MPC control-oriented model is a combination of the NN-based nonlinear steady-state model, trained from closed-loop historic data, and a linear dynamic model. However, the implementation of machine learning-based algorithms in building applications are recent.
ML algorithms are adapted in data-driven MPC design predominantly in two ways. First, a closed-loop MPC is used to generate the input-output data to be used as the training and testing data sets for an ML-based algorithm (\Cref{subsec_MPCtrainingRL}). Once trained offline, the ML algorithm replaces the MPC during run-time and thereby eliminating the receding horizon control optimization performed by the MPC at each control iteration. 
% Since the MPC generated data is used for off-line training, the control-oriented model is designed with rigorous detailing, ignoring the prolonged time of execution and computational complexity during the MPC optimization. The trained ML-based controller offers a time-efficient control in real-time. 
However, since the training data is generated using the MPC, the performance of the ML controller is still dependent on the control-oriented building model used by the MPC. 
% To overcome this model dependency, adaptive strategies may be implemented to periodically update the network parameters of the ML controller, based on the acquired information from the system during run-time.
% and implementing the controller in any other building requires major design alterations including the derivation of the control-oriented building model. 
% Research works following this trend are reviewed in \ref{subsec_MPCtrainingRL}.
Alternatively, another approach (\Cref{subsec_RLinMPC}) is to use ML-based algorithms  to design and train the control-oriented model offline, commonly using closed-loop control measurements from installed RBC or PID controller as the training data. The ML-based control-oriented model parameters are often updated periodically using various adaptive strategies. This control-oriented system model is subsequently used by the MPC for system output predictions during online optimization. 
% Adaptive strategies may also be implemented to update the ML based control-oriented
% A survey on this method is presented in \ref{subsec_RLinMPC}.

%%%%%%%%%%%%%%%%%%%%%%%%%%%%%%%%%%%%%%%%%%%%%%%%%%%%%%%%%%%%%%%%%%%%%%%%%%%%%%%%%%%%%%%%%%%%%%%%%%%%%%%%%%%%%%%%%%%%%%%%%%%

\subsection{MPC Generated Training Data for ML-Based Controller}
\label{subsec_MPCtrainingRL}

In \cite{DRGONA2018199} the authors propose an Approximate MPC where deep time delay neural networks (TDNN) and regression trees (RT)-based regression models are used as approximators for comfort and energy optimization of a multi-zone residential building. The training data is generated 
% by extracting time-varying system parameters and corresponding optimal control inputs from
using classical closed-loop MPC optimization profiles. Feature engineering is implemented to further reduce model complexity and implementation cost.
% by selecting the most significant features from the set of parameters using manual selection, principal component analysis (PCA) and selection of disturbance features based on model dynamics. However, discarding the less impactful parameters also imparts loss of physical meanings of the features in the derived model. 
% Finally, TDNN and RT-based controllers, once trained offline, replace the classical MPC during run-time. 
% As compared to the RBC baseline, the Approximate MPC with RT and TDNN improves 4\% and 12.8\% in energy saving, while the classical MPC 
% % used for generating the training data 
% achieves 16\% improvement. However,
TDNN-based Approximate MPC reduces the computation time approximately by a factor of 7 as compared to the classical MPC. 
% In their following publication, \textit{Drog\v{n}a et al.} proposed 
A real-time implementation of a deep-learning-based policy approximator for MPC (DL-MPC) is later presented by \cite{drgona2019stripping} in an office building located in Hasselt, Belgium. 
% Classical optimization-based MPC (OB-MPC) trajectories are used for DL-MPC optimal policy training. RBC is used as a common baseline for both DL-MPC and OB-MPC. DL-MPC is reported to achieve close to 100\% comfort satisfaction with a 4\% increase in energy cost as compared to OB-MPC. Here also, DL-MPC is roughly 50,000 times faster than OB-MPC in terms of CPU time and uses 638-times less memory. This significantly reduces the software deployment cost. 
A deep neural network (DNN)-based microcontroller is presented in \cite{Karg_2018}. 
% The microcontroller is trained offline using data from 500 different runs of a classical MPC. 
The DNN-based controller has a fast computation time and achieves an `almost globally optimal solution' of a mixed-integer quadratic program (MIQP).
% used to formulate indoor temperature and energy optimization problems in a smart house with a single thermal zone.

%%%%%%%%%%%%%%%%%%%%%%%%%%%%%%%%%%%%%%%%%%%%%%%%%%%%%%%%%%%%%%%%%%%%%%%%%%%%%%%%%%%%%%%%%%%%%%%%%%%%%%%%%%%%%%%%%%%%%%%%%%%

\subsection{Data-driven Predictive Model for MPC}
\label{subsec_RLinMPC}

An artificial neural network (ANN)-based MPC driven HVAC control system is presented \cite{FERREIRA2012238}. Radial basic function-based neural networks (RBF-ANN), 
% representing the predicted mean vote (PMV) index function driven prediction model for comfort, 
are identified by a multi-objective genetic algorithm (GA). The ANN models achieve a suitable trade-off between the accuracy of predicted mean vote (PMV) approximation for comfort and the execution time of the MPC. 
The discrete MPC uses a branch and bound technique to find the optimal control signal. In \cite{AFRAM201796} the authors present a list of publications, between 2010 to 2016, on the ANN-aided MPC approach in various BEMS applications.
% focusing on control of air handling unit and variable air volume terminals, air conditioning system with variable speed compressor for comfort management, HVAC controls for office and school buildings etc. 
In this article, the authors also implemented an ANN-based MPC for a residential building located in Vaughan, Ontario, Canada. The best network after multiple iterations (BNMI) approach has been used to determine the appropriate ANN-driven predictive models for HVAC components. The MPC-driven controller is used in a supervisory level controlling setpoints of local PID controllers. 
% Energy and cost optimization by load shifting problem has been addressed in \cite{REYNOLDS2018729} using ANN-based energy consumption and indoor temperature forecast model. A GA-based day-ahead optimal temperature setpoint scheduling approach with an ANN forecast model is compared with the same ANN-aided MPC strategy. In reference to the pre-scheduled baseline, both the day-ahead GA-ANN controller and ANN-based MPC show roughly a 20\% to 30\% improvement with GA-ANN performing slightly better than the MPC. However, as mentioned by the authors, 100\% ANN forecast accuracy is assumed in the testbed experiment. In a more practical situation, an ANN-based MPC might better handle the disturbances imposed by the forecast errors. 
\cite{Zhang2019} introduces safety aware exploration using model-based deep RL for MPC control-oriented model identification. MPC minimizes energy cost and zone temperature violation using a random-sampling shooting method. The time constraint in real-time implementation is tackled by training an auxiliary policy network that imitates MPC outputs. The proposed method achieves 17.2\%$\sim$21.8\% reduction in total energy consumption along with 10$\times$ reduction in total required training steps as compared to model-free RL using proximal policy optimization (PPO). A novel model-based deep RL algorithm, namely MB$^2$C, is presented in \cite{Ding2020}, which improves the results presented in \cite{Zhang2019}. Here, the authors proposed a Model Predictive Path Integral (MPPI) based control approach instead of a random-sampling shooting method \cite{Zhang2019} for the MPC optimization since the latter is not efficient in identifying the best action as the randomly chosen batch of action trajectories may not include the best action sequence. The control-oriented building model, designed for predicting changes in the system states, is identified as an ensemble of multiple adaptive environment-conditioned neural networks (ENNs) which take into account environmental disturbance inputs along with system state and action as network inputs. A weighted ensemble learning approach is adopted to train these ENN models where each model is initialized using random initial network parameters from different batches. MB$^2$C achieves 8.23\% more energy savings as compared to the MBRL controller in \cite{Zhang2019} while maintaining similar comfort. It also achieves higher data efficiency and faster convergence as compared to a model free PPO-based RL baseline controller as well as the MBRL controller in \cite{Zhang2019}. 

A continuous building model adaptation scheme using an ANN-based MPC control-oriented model is proposed in \cite{YANG2020115147}. 
% The ANN model is initially trained offline using BAS data. Later on, 
The adaptive ANN model is updated in each control iteration using the closed-loop MPC data. The ANN-aided MPC has achieved a 58.5\% cooling energy reduction along with a 36.7\% reduction in electricity consumption while maintaining a comfortable indoor environment. In \cite{Eini_2020}, ANN-based MPC is used for lighting and thermal comfort optimization. A nonlinear autoregressive exogenous model with parallel architecture (NARX) is used to train the networks that estimate the PMV-based comfort specifications, environmental conditions and power consumption. 
% ANN-based MPC improves visual comfort by 45.2\% with 7.25\% improvement in PPD (predicted percentage of dissatisfied) for comfort. 
An input convex neural network (ICNN) quasi-convex MPC is proposed in \cite{bunning2020input} to ensure input-output convexity of ANN mapping for the control-oriented model. However, ICNN generally only guarantees one-step prediction convexity. In this work, an extension is proposed for multi-shot multi-step prediction convexity with a feed-forward network. 
% A temporal-sequential ANN (TS-ANN) based occupancy pattern prediction model is proposed in \cite{jin2020data} to improve MPC performance for energy optimization through efficient lighting control. 

% An experimental implementation of energy optimization and zone-based comfort management with a dedicated morning start module has been presented. 
A random forest (RF) based control-oriented model for MPC is used in \cite{HILLIARD2017326}. A 29\% reduction in HVAC electrical energy consumption and a 63\% reduction in thermal energy consumption is achieved. % reported during a period of four months between August to November. 
% Regression trees and forests-based predictive models for MPC for building thermal comfort and energy optimization are also presented in  \cite{Jain_2017CDC, SMARRA20181252}. 
In \cite{Jain_2017CDC} the authors propose data-predictive control (DPC) 
where
% bilinear control-oriented building model in classical 
MPC uses regression trees (RT) for prediction. Each RT represents an affine function, that relates the system outputs to the control inputs from the training data, associated with a prediction time step. 
% Each leaf of the regression trees is associated with an affine function that relates the system outputs to the control inputs from the training data. The MPC solves a conventional constrained optimization problem to generate the optimal control signal during run-time, where the constraints are imposed by the affine functions generated by the RTs. 
To alleviate overfitting and high variance of the RTs, an ensemble learning approach is adopted replacing RTs with RFs for each timestep. A variation of the similar approach is presented in \cite{SMARRA2018199}, where state-space switched affine dynamical linear time-invariant (LTI) models are identified instead of the affine static prediction 
% models as mentioned earlier. 
The switched affine state-space models take into account the internal state evolution. 
Data predictive control based on RFs with affine functions and convex optimization problem for BEMS is presented in \cite{BUNNING2020109792}. The high dimensionality of the affine function coefficient fitting process is simplified by 
% limiting the number of fitted coefficients.
choosing only two fitted coefficients, implying all past control inputs have similar effects on the state as that of the current control input. Experiment results show better model performance even though the assumption is less realistic as explained by the authors.  

\cite{COTRUFO2020109563} uses Gaussian process regression (GPR) based MPC control-oriented models to predict heating demand, electric baseload and natural gas consumption for a gas consumption minimization problem and implemented them at Canmet ENERGY-Varennes, an NRCAN research facility at Qu\'{e}bec, Canada. 
% A comparison of five machine learning techniques, namely, ANN, GPR with squared exponential kernel function, support vector machine (SVM), decision trees and RF are presented. GPR, having the best accuracy, is chosen for the model identification. 
About 22\% reduction in natural gas consumption and greenhouse emission is reported, along with 4.3\% reduction in the net building heating demand as compared to the current building control operation.

\section{Review of RL in BEMS}
\label{sec_RL_survey}

By feeding the predictive information into the dynamic model,  MPC has shown its possibilities to improve the control performance of BEMS. In this section, recent advances of RL in building control are reviewed, and they show that RL can produce good results through interactions between the agents and building environments . 
%and they show that RL can get good results by modelling the dynamic process flexibly. 

\subsection{Classical RL Methods in BEMS}
\label{subsec_ClassicalRL}

Classical RL methods such as Q-learning and SARSA are found to dominate the learning algorithms in building control before 2017 \cite{9320042}.
%The author in \cite{7042790} formulated a fully automated 
A pioneering work in RL-based energy management by Google DeepMind, which was developed using the RL method, was shown to decrease the electricity bill for cooling a data center by approximately 40\%. The home energy management problem is formulated as an MDP and then solved with a Q-learning algorithm in~\cite{7042790}. A method that combines the tree-like MDP and  SARSA algorithm was proposed for appliance scheduling and compared with a Q-learning in~\cite{7917970}. Experimental results show that both SARSA and Q-learning obtain a similar schedule for a finite number of appliances over a 24-hour horizon, but the schedule is arrived at much faster using the variation of SARSA. The authors in \cite{al2016demand} compare the value-based approach with the Actor-Critic method for domestic hot water control and find Q-learning performs better.

To combine the users' feedback into the control process and smooth the power consumption profile, the authors in \cite{9000577} propose an algorithm directly integrating user feedback into its control logic using fuzzy reasoning as reward functions. Then, Q-learning is used to make optimal decisions to schedule the operation of smart home appliances.

However, classical RL methods suffer severely from the curse of dimensionality when using historical data or applied to continuous operations in building control. There has been some effort to reduce the dimension in training data.  For instance, only states of the previous two-time steps are considered in the critic network, markedly decreasing the number of inputs \cite{fuselli2013action}.  The authors in \cite{ruelens2015learning} use a deep neural network-based dimension reduction technique, then compress the previous ten indoor temperatures and control signals into six hidden states. In their following work \cite{ruelens2016residential}, the predicted states are used to help improve the performance of the RL controller as MPC, and they find that including weather forecasts as states could improve the performance by 27\%.

\subsection{Deep RL Methods in BEMS}
\label{subsec_DeepRL}

In continuous control and discrete control problems which have large action spaces and state spaces, the curse of dimensionality hinders the implementation of RL in practice. So, deep neural networks have been widely used recently as function approximators to represent policies, rewards and value functions in building control to scale up the RL algorithms. The deep Q-learning has been adopted to optimize the operation cost or energy consumption of a single building HVAC system through EnergyPlus tool in \cite{wei2017deep}, where the system determines discrete air flow rate based on time, zone temperature and environment disturbances. The authors in \cite{vazquez2017balancing} substitute the Q-table with an ANN that maps current and target temperature directly to their Q-values, allowing the controller to work with continuous states and actions, and also to speed up the learning process, and experiments results demonstrate that the deep RL-based algorithm is more effective in energy cost reduction compared with the traditional rule-based approach. The authors in \cite{kazmi2018gigawatt} implement the deep RL hot water controllers in 32 Dutch houses by taking into account occupant interaction and hot water system dynamics.  They find that, compared with the fixed schedule or fixed setpoint control, the RL controller reduces energy consumption by almost 20\% while maintaining occupant comfort.  %Then the Deep RL techniques have been adopted in water heater system in building control \cite{ruelens2019direct}.
 
 Except for controlling single device in building, there also exist some methods consider the control operations of heterogeneous home appliances and distributed energy resources according to the consumer’s comfort and preferences. The authors in \cite{lee2020energy} propose a two-level hierarchical deep RL-based energy management framework for HEMS to handle the interdependent operation between the home appliances at the first level and the energy storage system (ESS)/EV at the second level. Then the optimal policy for charging and discharging actions of the ESS and EV is independently determined considering the energy consumption schedule of aggregated home appliances. The approach employs an actor-critic method where the controllable home appliances are scheduled at the first level. The energy-saving system and EVs are scheduled at the second level to cover the aggregated washing machine (WM) and air conditioner (AC) loads, which are calculated at the first level along with the fixed load of the uncontrollable appliances. A DQN-based home energy management that considers both the appliances scheduling and EV charging scheduling is presented in~\cite{di2018optimizing}. To deal with both discrete and continuous actions to jointly optimize the schedules of all kinds of appliances, a deep RL approach based on trust region policy optimization is proposed in \cite{9022934}. The approach considers three kinds of appliances including deferrable appliances, regulatable appliances, and critical appliances in the simulation model and directly learns from raw observation data of the appliance states, real-time electricity price, and outdoor temperature. Apart from the BEMS in a residential home, there are many existing works on building control with consideration of commercial building. 
The authors in \cite{ding2019octopus}  propose a deep RL-based framework for efficiently controlling four-building energy subsystems so that the total energy consumed by all subsystems can be minimized while still maintaining user comfort.
 
 Another way to deal with multiple appliances or multiple buildings control problems is to use multi-agent  (MA) RL, where each agent corresponded to various home appliances/buildings can communicate with each other. In classical RL algorithms and deep RL works, they often investigate energy policies for household appliances under same environments and reward setting, thereby restricting the algorithm’s effectiveness and generalization in real scenarios. MARL requires setting several agents, where different types of household appliances represent different agents with their actions and rewards. The authors in \cite{lee2019reinforcement} propose a methods for the optimal scheduling of different household appliances to optimize energy utilization. The authors in \cite{nagarathinam2020marco} propose a 
MARL algorithm to minimize HVAC energy consumption without sacrificing user comfort by adjusting both the building and chiller set-points. To speed up the training process,
they use transfer learning in which the agents are trained on sub-sets of HVAC systems and the learned network weights are used to initialize the
multiple agents. Furthermore,  an hour-ahead DRL algorithm for HEMS based on multi-agent RL is proposed in \cite{8681422}, which optimizes both shiftable appliances and AC considering the uncertainty in future prices. The consumption scheduling problem is first formulated as a finite MDP (FMDP) with discrete time steps. Then, the FMDP is utilized to model the hour-ahead energy consumption scheduling problem to minimize the electricity bill, as well as DR, including dissatisfaction. The authors in \cite{sun2020continuous} focus on the large-scale HEMS optimization problem for smart homes and propose a collective MARL algorithm with continuous action space to achieve flexible and precise control. Apart from 
MARL, multi-objective learning also received research attention considering different control objectives in BEMS. A multi-objective algorithm is proposed based on human appliances interaction, which considers scheduling in the context of energy consumption and discomfort level of the home user \cite{diyan2020multi}.

However, some works also point out that it is impractical to let the deep RL agent explore the state space fully in a real building environment because an unacceptably high cost may be incurred. Plus, it may take a long time for the deep RL agent to learn an optimal policy if trained in a real-world environment. To reduce the dependency on a real building environment, many model-based deep RL control methods have been developed and some research has tried to incorporate the domain knowledge into the training process or suggests using the MPC method to enhance the understanding of the dynamic process. 
The authors in \cite{zhang2019whole} use the observed data in EnergyPlus to develop a building energy model, and then use the model as the environment simulator to train the deep RL agent off-line based on the A3C algorithm. In this way, the deep RL agent’s potentially harmful exploration in real-world HVAC is limited. In \cite{zou2020towards}, LSTM is used to build the environment model using historical data, in which the inputs of LSTM models are the current state and action, while their outputs are the next state and reward. Then the agent is trained using DDPG. 
The authors in \cite{chen2019gnu} encode the domain knowledge on planning and linear system dynamics into the RL controller by differentiable MPC policy. The system uses offline pre-training by imitating the existing controller and online learning to interact with the environment and update its policy. The results show that it can save 16.7\% of cooling demand compared to the manual set-point controller. 
What also deserves to be mentioned is that training an RL controller is data and time-demanding. To accelerate the training process, the approach proposed in \cite{9000577} works with a single agent and uses a reduced number of state-action pairs as rewards functions. The authors in \cite{zhou2019artificial} uses fuzzy rules to discretize continuous states-actions and to reduce dimensions in a smart residential community. The approach proposed in \cite{yoon2019performance} uses Gaussian Process Regression (GPR) to compress six states into two.

\section{Challenges and Considerations in the Selection of Control Methods}
\label{sec_challenges}

This section focuses on the prevalent challenges associated with the real-time implementation of the MPC and RL-driven control methods in different BEMS applications. Also, some considerations on the potential choices of control methods to address these challenges are discussed. 

%%%%%%%%%%%%%%%%%%%%%%%%%%%%%%%%%%%%%%%%%%%%%%%%%%%%%%%%%%%%%%%%%%%%%%%%%%%%%%%%%%%%%%%%%%%%%%%%%%%%%%%%%%%%%%%%%%%%%%%%%%%%%%%%%%%%%%%%%%%%%%%%%%%%%%%%%%%%%%%%

\subsection{Challenges of MPC Implementation}
\label{subsec_MPC_challenges}

The main drawback of MPC-based BEMS control is the restrictions imposed by the model-based prediction strategies for control signal optimization. The limitations are twofold. First, while a sufficiently accurate building model is crucial for efficient performance of an MPC, the design and calibration of such model is complex. Presently, with the development of smart communities, shared energy storage and an increasing number of EVs participating in the grid-based demand-side management, the complexity of the BEMS is growing rapidly. Secondly, the repeated optimization at every time step ensures robustness 
% against abrupt changes in the external disturbances, 
but demands serious run-time computational capacity leading to increased initial investment and operational cost. 

% \textit{Cigler et al.} effectively summarizes four challenges of the classical MPC in building control. These are related to-- 
 
%  \renewcommand{\labelenumi}{\alph{enumi}.}
%  \begin{enumerate}
%      \item mathematical modelling accuracy of the predictive control oriented model.
%      \item controller tractability for large multi-zone buildings.
%      \item memory and time constraints for hardware and software implementation.
%      \item data availability and its processing needed for the design of the control-oriented model.
%  \end{enumerate}
  
 Although the design specifications of modern buildings are more accessible nowadays, old structure retrofits pose a hindrance in model identification. Eventually, this also limits the possibilities of fast adaptation of classical MPC, as no building is identical to the other. It is worth mentioning that the data-driven MPCs, reviewed in \ref{subsec_MPCtrainingRL}, also suffer from a similar drawback. Even though they perform impressively to resolve the time-expensive nature of the classical MPC optimization routines, the data-driven MPCs are restricted to specific buildings when trained on data generated by the `teacher' MPCs. Though MPC is theoretically capable of handling multi-objective optimization goals simultaneously, the time and hardware constraints may make it practically infeasible for real-time implementation. However, unlike purely RL-based control, a classical MPC ensures reliability from the start of its operation, given a sufficiently accurate traditional physics-based control-oriented model or a grey box model based on thermal resistance-capacitance. A predictive model-based RL approach may offer a promising path forward to achieve an agreeable trade-off between the implementation time and robustness. 
 
% Unlike purely data-driven prediction approaches, traditional physics-based models or thermal resistance-capacitance (RC) based grey box models for MPC, safeguard it from being too dependant on the sensor-based measurement system. Close-loop data from sensors have implicit biases based on the measurement accuracy or positioning at different locations in the thermal zones, which severely affect the accuracy of the data-driven prediction models. Consequently, a model-free data-driven approach has little to do to recover from these inherent setbacks.

%  However, a more accurate model might not necessarily lead to better control \cite{Chen2019}. A twin building with the same architecture may need its control-oriented model based on its location, positioning etc., since based on these factors the external inputs, such as exposure to the solar radiation on the building fa\c{c}des, can vary widely.  

%%%%%%%%%%%%%%%%%%%%%%%%%%%%%%%%%%%%%%%%%%%%%%%%%%%%%%%%%%%%%%%%%%%%%%%%%%%%%%%%%%%%%%%%%%%%%%%%%%%%%%%%%%%%%%%%%%%%%%%%%%%%%%%%%%%%%%%%%%%%%%%%%%%%%%%%%%%%%%%%

\subsection{ Challenges of RL Implementation}

Even though RL-based BEMS management has recently attracted increased research interest , %the RL controller is still in the research and development stage and only a few RL controllers were implemented and tested in an actual building. So 
a lot of research challenges and questions are still open for further research.

The foremost challenge is data efficiency. Most of the current approaches rely heavily on accurate simulator design and a large amount of training data. %that is representative of real-world scenarios. 
However, those data are hard to collect let alone be used in reality. Taking recently published results as examples, an RL agent may need 5 million interaction steps (47.5 years in simulation) to achieve the same performance as a feedback controller on an HVAC system \cite{zhang2018practical}. Furthermore, while one can train an RL agent in simulation, it is not cost-effective to create a model for each home. So how to improve data efficiency and use data-efficient RL algorithms in the smart grid is worthy of further investigation. %Besides, implementing transfer learning is a potential solution so that controllers trained by a small number of smart homes could be generalized and used for other houses.

The second challenge is the safety requirement of RL used in building control, which is mostly neglected by the current algorithm design. Since the environment states cannot all be covered in the training phase, the learned exploratory control policy may bring the system into potentially dangerous states, such as power surge which could crash the whole power system. There exist some works using a backup controller to enhance control safety when there could be dangerous states \cite{ruelens2016reinforcement,de2017using}, and others use the safe boundary to limit the agents' action \cite{chen2019gnu}. However, none of them could give a general idea on how to safely train and implement controllers in real buildings.
%There has been some research that focuses on safe RL using safe constraints or chance-constrained safety set to solve such problem theoretically \cite{alshiekh2018safe,berkenkamp2017safe}. It is promising and pressing to deal with more robust and safe reinforcement learning when we plan to apply such RL based control strategy to BEMS.

Moreover, the vast majority of current studies included in this survey are in simulation only. Most of the studies of RL in building control are not easily reproducible, and the benchmarks for results are not unified. Some researchers use fixed setpoints while others use basic RL as baselines, which may result in unfair comparison.  The OpenAI Gym environment, CityLearn, was developed and open-sourced, aiming for the easy implementation of reinforcement learning controllers in a multi-agent demand response setting. Further, more efforts are needed to better integrate the computation platforms and create more general, data-rich and inter-operable virtual test benches.
 
%Plus, the change in people's behaviour patterns and human impacts on the control of HEMS has rarely been discussed, such as occupancy and equipment usage patterns. Wei et al. [62] \cite{wei2020deep} design a DRL-based recommender system in commercial buildings, which can learn actions with high energy saving potential and distribute recommendations to occupants. Based on the feedback from occupants, better recommendations can be learned.

%%Markovian Property represents the behaviour that future states purely depend on the current states, which, unfortunately, does not hold for building thermal dynamics, because of the thermal mass. To solve this problem, historical states need to be included in the MDP, most controllers using RL but cannot guarantee that the Markovian Property holds. 

%%%%%%%%%%%%%%%%%%%%%%%%%%%%%%%%%%%%%%%%%%%%%%%%%%%%%%%%%%%%%%%%%%%%%%%%%%%%%%%%%%%%%%%%%%%%%%%%%%%%%%%%%%%%%%%%%%%%%%%%%%%%%%%%%%%%%%%%%%%%%%%%%%%%%%%%%%%%%%%%

\subsection{Considerations for Choice of Control Strategies}

Although many challenges exist, data-driven MPC and RL-based methods are believed to be the most powerful strategies to handle decision-making problems in BEMS. Based on specific control objectives, both strategies can be well designed, selected and even combined to get the desired control performance. The key features of the two methods can be summarized as follows:

\begin{enumerate}
    \item \textbf{Length of time horizon:} MPC optimizes finite-length trajectories based on a pre-specified or learned system model. RL-based algorithms can find an infinite horizon optimal policy under unknown dynamics of the system, based on interactions and external cost signals (where dynamics could also be learned in model-based RL).
    
    \item \textbf{Data used for control design:} RL is based on experience, which is used to reduce the need for iterative methods. Moreover, depending on the formulation of the problem and the richness of experience data, the chances of convergence are high. MPC, on the other hand, integrates forecast information and considers future disturbances to handle multiple constraints and objectives. However, it is limited by the need for accurate models. This is especially challenging because buildings are heterogeneous.
    
    \item \textbf{Control strategy:} RL explicitly considers the whole problem of a goal-directed agent interacting with an uncertain environment. However, the flexibility of the RL framework comes at the cost of increased sample complexity. Furthermore, while these results show that RL agents can be trained successfully in simulation, high-fidelity models are generally not available for building thermal zones, whereas MPC treats modelling and planning as two separate tasks. The quality of the model is evaluated by criteria such as prediction error, which may affect control performance. Data-driven MPC approaches adapt their models online, based on observations from the system. This allows the controller to improve over time, given limited prior knowledge of the system. However, the method relies on extensive offline computations. 
    
    \item \textbf{Overlap between MPC and RL:} In model-based RL, the model could be a data-driven model, such as a deep neural network, or a physics-based model, such as thermal resistance–thermal capacity model. In this regard, a model-based RL is similar to the MPC technique. Moreover, the prediction information of systems could also help the RL to correct transition probabilities and thus resulting in more accurate control.
\end{enumerate}

In summary, a characterization of the current research contributions in data-driven MPC and RL-based methods in BEMS can be made based on computational complexity, data-efficiency, safety and robust adaptability. The ML-based controllers trained using classical MPC (\Cref{subsec_MPCtrainingRL}) reduce on-line computational time as compared to the classical MPC, by eliminating the necessity of on-line control optimization at each time step, but they still need to model the system dynamics accurately. Learning the optimal policy by testing new policies and evaluating the outcomes, classical RL (\Cref{subsec_ClassicalRL}) and deep RL (\Cref{subsec_DeepRL})-based algorithms applied in BEMS rely less on precise modelling while having low data efficiency and reliability in general. MPC with a data-driven predictive model (\Cref{subsec_RLinMPC}), on the other hand, preserves the reliability ensured by the classical MPC while simplifying the predictive model identification process. Nonetheless, it appears that neither RL-based nor data-driven MPC algorithms have yet successfully integrated sufficient adaptability to be commercially implemented at a wide scale for BEMS in the heterogeneous building stock.

\section{Conclusion}
\label{sec_conclusion}

% In this article a review on the recent developments in the data-driven MPC and RL-based control algorithms for BEMS is presented. The challenges concerning the different approaches have been discussed. Finally, considerations on the choice of control methods have been presented based on an application perspective. \cref{tab_procon} may be used as a quick guide for selecting preliminary control strategies based on given control objectives. 
In this paper, a compact survey of recent developments in data-driven MPC and RL-based control algorithms for BEMS has been presented. Data-driven MPC faces the challenges of design complexity and time-consuming computations, while RL-based methods face the data-efficiency, safety and robust adaptability problems. Considerations for the choice of control methods in real-world applications have been presented. Combining the classical MPC with RL-based prediction approaches appears to offer a suitable trade-off between reliability and practicality of implementation. For MPC, the data-driven approaches have contributed to improving computational complexity, time constraints, adaptability and simplification of the control-oriented model design which is an essential part of predictive control. For RL-based BEMS control, MPC can contribute to ensuring safety and robustness. Relatively simpler data-driven predictive models combined with robust control strategies seem to chart a reasonable path forward in order to achieve desired time-efficiency, reliability and adaptability for real-time building energy management. 
\bibliographystyle{IEEEtran}
\bibliography{ref_survey_rl,ref_intro,ref_challg,ref_survey_ss,ref_background}

% \fi
\IEEEpeerreviewmaketitle

\end{document}